\documentclass[12pt]{article}
\usepackage{latexsym}
\usepackage{amsbsy}
\usepackage{graphicx}

\newsavebox{\sboxpubnumber}
\newsavebox{\sboxpubdate}
\newcommand{\pubdate}[1]{\begin{lrbox}{\sboxpubdate}{#1}\end{lrbox}}

\newcommand{\Title}[1]{\begin{center} {\Large #1 } \end{center}}
\newcommand{\Author}[1]{\begin{center}{ \sc #1} \end{center}}
\newcommand{\Address}[1]{\begin{center}{ \it #1} \end{center}}
\newcommand{\andauth}{\begin{center}{and} \end{center}}

\newenvironment{Abstract}{\begin{quotation}  }{\end{quotation}}
\newenvironment{Presented}{\begin{quotation} \begin{center}
             PRESENTED AT\end{center}\bigskip
      \begin{center}\begin{large}}{\end{large}\end{center}
      \end{quotation}}
\newcommand{\Acknowledgements}{\bigskip  \bigskip \begin{center} \begin{large}
             \bf ACKNOWLEDGEMENTS \end{large}\end{center}}

\newcommand{\vrh}{\hat{\vr}}
\newcommand{\vrhd}{\hat{\vr}^{\dagger}}

\newcommand{\pih}{\hat{\pi}}
\newcommand{\pihd}{\hat{\pi}^{\dagger}}

\newcommand{\al}{\alpha}
\newcommand{\bt}{\beta}

\newcommand{\dl}{\delta}

\newcommand{\lm}{\lambda}
\newcommand{\rh}{\rho}

\newcommand{\vr}{\varphi}

\newcommand{\om}{\omega}

\newcommand{\half}{\frac{1}{2}}
\newcommand{\quart}{\frac{1}{4}}

\newcommand{\eela}[1]{\label{#1}\end{equation}}
\newcommand{\eeala}[1]{\label{#1}\end{eqnarray}}
\newcommand{\ra}{\rightarrow}

\newcommand{\be}{\begin{equation}}
\newcommand{\ee}{\end{equation}}
\newcommand{\bea}{\begin{eqnarray}}
\newcommand{\eea}{\end{eqnarray}}
\begin{document}

\begin{titlepage}
\pubdate{November 2001}                    

\vfill
\Title{Initial conditions for simulated `tachyonic preheating' and the
Hartree ensemble approximation}
\vfill
\Author{Jan Smit}
\Address{ }
\vfill
\andauth
\vfill
\Author{Jeroen C.\ Vink and Mischa Sall\'e}
\Address{Institute for Theoretical Physics, University of Amsterdam, \\
         Valckenierstraat 65, 1018 XE Amsterdam, the Netherlands}
\vfill
\begin{Abstract}
In numerical simulations studying preheating 
in the classical approximation there is the problem how to derive
the classical initial conditions from the quantum vacuum fluctuations.
In past treatments, the initial conditions often put an
energy density into the classical field of order of the cutoff,
leading to a divergent temperature after thermalization.
We suggest a solution to the problem which follows naturally
from a Hartree ensemble approximation, introduced recently as an improvement
over the standard Hartree approximation.
We study the effects on particle numbers of the various treatments, 
within the context of `tachyonic preheating' in 1+1 dimensional
$\varphi^4$ theory.
\end{Abstract}
\vfill
\begin{Presented}
    COSMO-01 \\
    Rovaniemi, Finland, \\
    August 29 -- September 4, 2001
\end{Presented}
\vfill
\end{titlepage}
\def\thefootnote{\fnsymbol{footnote}}
\setcounter{footnote}{0}

\section{Introduction}

Numerical simulations of quantum field dynamics in real time
are currently carried out by first making approximations. 
For systems out of equilibrium the 
classical approximation can be quite useful, for example in studying
reheating after inflation \cite{Khlebnikov:1996wr}-\cite{Rajantie:2000fd}. 
In addition there are Hartree
(gaussian), Dyson-Schwinger and 2PI-functional approximations, 
possibly supported by large $N$ considerations.
These have the advantage of being formulated in the 
quantum theory where potential divergencies are taken care of
by renormalization. 

Even if the system is homogeneous it may be important to allow for
inhomogeneous realizations in order to capture non-perturbative effects.
Recently we have formulated a Hartree ensemble approximation 
\cite{Salle:2000hd,Salle:2000jb}, in which an overall homogeneous 
density matrix is represented as 
an ensemble of typically inhomogeneous initial configurations,
which are then evolved in time using the Hartree approximation. 
With this method we found that the poor thermalization properties
of the homogeneous Hartree approximation are much improved.

Here we shall describe an application of this method
to the problem of spinodal or tachyonic instability. This
process has been put forward as an efficient mechanism for preheating.
In ref.\ \cite{Felder:2001kt,Felder:2000hj} 
it was studied numerically, using the classical approximation
with initial values based on quantum fluctuations. However, 
these initial conditions (of the type also used in the other studies in
\cite{Khlebnikov:1996wr}-\cite{Rajantie:2000fd})
lead to problems: the final temperature is of order of the cutoff. 
Below we propose improved initial conditions for the classical approximation
which do not have this 
problem and we compare with results obtained using the previously
employed initial conditions, in classical and Hartree dynamics.

We also draw the attention to a fruitful
definition of particle numbers and energies
for systems out of equilibrium, which was introduced in \cite{Salle:2000hd},
and for the fermionic case in \cite{Aarts:1999zn}.

\section{Hartree ensemble approximation}
Our tests are carried out in in 1+1 dimensional $\vr^4$ theory. 
The quantum hamiltonian is given by
\be
 \hat H = \int dx\, \left(\half \hat{\pi}^2 +\half (\nabla \hat\vr)^2
   + \half \mu^2 \hat{\vr}^2 + \quart \lm \hat{\vr}^4\right),
\ee
from which follow the Heisenberg operator equations
\be
\dot{\hat{\vr}} = \hat{ \pi}, \;\;\;\;
\dot{\hat{\pi}} = (\nabla^2 - \mu^2)\hat{\vr} - \lm \hat{\vr}^3.
\ee
We discretize the field on a space-time lattice, with lattice spacings
$a$ and $a_t$, such that $\nabla \vr \ra (\vr_{x+a} - \vr_{x})/a$,
$\dot{\vr} \ra [\vr_x(t+a_t) - \vr_x(t) ]/a_t$. The number of spatial lattice
sites is $N$, the volume $L = Na$ and we use periodic boundary conditions.
The Hartree approximation implies that the fields can be expanded in terms
of mode functions
\be
\hat \vr_x = \vr_x + 
\sum_{\al} \left( f^\al_x\; \hat{b}_\al +f^{\al *}_x \;\hat{b}^{\dagger}_\al
\right),
\;\;\;\;
\hat{\pi}_x = \pi_x + \sum_\al \left(\dot{f}^\al_x \;\hat{b}_\al
+ \dot{f}^{\al *}_x \;\hat{b}^{\dagger}_\al\right),
\ee
in which $\vr_x$ and $\pi_x = \dot\vr_x$ are the mean fields.
The creation and annihilation operators
$\hat{b}_\al^\dagger$, $\hat{b}_\al$ are time-{\em in}dependent, with 
$[\hat{b}_\al, \hat{b}_\bt^\dagger] = \dl_{\al \bt}$, while the mode functions
$f_x^{\al}$ are time-dependent. Furthermore, the density operator (which is
time-independent in the Heisenberg picture) is gaussian, such that it is
completely specified by giving the initial values of the one and two-point
functions, 
$\langle\hat\vr_x\rangle = \vr_x$, $\langle\hat\pi_x\rangle = \pi_x$,
and $\langle\vrh_x\vrh_y\rangle$, $\langle\vrh_x\pih_y + \pih_y\vrh_x\rangle$,
$\langle\pih_x\pih_y\rangle$. With a suitable choice of initial mode functions
these can be rephrased as
$\langle \hat b_{\al}^{\dagger}\hat b_{\bt}\rangle = n^0_{\al}\dl_{\al\bt}$,
with $\langle \hat b_{\al}\rangle = \langle\hat b_{\al}^{\dagger}\rangle = 
\langle \hat b_{\al} \hat b_{\bt}\rangle =
\langle \hat b_{\al}^{\dagger} \hat b_{\bt}^{\dagger}\rangle = 0$.
In case the gaussian state is pure, the $n_{\al}^0 =0$, and the Hartree
equations for the mean field and mode functions take the form of a 
set of coupled non-linear partial differential equations,
\bea
   \ddot{\vr}_x &=& \nabla^2\vr_x - \left(\mu^2 + \lm \vr_x^2 +
             3\lm\sum_{\al} f_x^{\al} f_x^{\al *} \right)\vr_x ,
 \nonumber \\
   \ddot{f}_x^{\bt} &=& \nabla^2 f_x^\bt - \left( \mu^2 + 3\lm\vr_x^2 +
      3\lm\sum_{\al} f_x^{\al} f_x^{\al *} \right) f_x^\bt.
                   \label{EOMH}
\eea
In 1+1 dimensions the mode sum is only logarithmically divergent as the
lattice spacing $a\to 0$, which is taken care of by mass renormalization.
In the following we assume the divergence to be subtracted from the mode
sum, such that $\mu^2$ is a renormalized mass parameter, with particle
mass $m^2 = \mu^2$ (symmetric phase) and $m^2 = -2\mu^2$ (broken phase).

In the {\em Hartree ensemble approximation} the density operator $\hat\rh$ is
written as a superposition in terms 
of gaussian coherent pure states $|i\rangle$,
\be
\hat\rho = \sum_i p_i |i\rangle\langle i|,
\label{rhoensemble}
\ee
and the Hartree approximation is applied to each state $|i\rangle$
individually.
Note that $\hat\rh$ does not have to be gaussian or pure-state.
Given $\hat\rh$ and a choice of the set of coherent states $|i\rangle$, 
the probabilities $p_i$ are uniquely defined in simple cases,
e.g.\ a free thermal ensemble \cite{Salle:2000hd}. 
However, we shall use (\ref{rhoensemble}) to formulate approximations to
the true $\hat\rh$, such that the initial values of the one and two-point
functions are reproduced. We see this as a sort of coarse graining:
not all details of the true $\hat\rh$ are kept, only those referring to the 
important low momentum
modes of the one and two point functions $\langle\hat\vr_x\rangle$, 
\ldots, $\langle\pih_x\pih_y\rangle$. 

Even if $\vr_x = \langle\vrh_x\rangle = 
\sum_i p_i \langle i|\vrh_x|i\rangle$
is homogeneous (independent of $x$), the individual 
$\vr_x^{(i)} = \langle i|\vrh_x|i\rangle$ are typically inhomogeneous.
This is important for thermalization, because the modes interact non-linearly
with the inhomogeneous $\vr_x^{(i)}$: the particles can scatter. 

\section{Particle number}
One needs a definition of particle number $n_k$ and frequency $\om_k$
which is reasonably intuitive and robust, 
such that it can be applied to systems out of equilibrium. Some
coarse graining will have to be involved and the following definition
takes this to the extreme of using an average over the whole volume:
we define $n_k$ and $\om_k$ at each time by the equations
\be
\langle \vrh_k\vrhd_k\rangle_{\rm conn} =  
\left(n_k + \half\right)\, \frac{1}{\om_k},
\;\;\;\;
\langle\pih_k\pihd_k\rangle_{\rm conn} =
 \left(n_k + \half\right)\, \om_k,
\label{defnk}
\ee
where $\vrh_k = \sum_x e^{-ikx}\, \vrh_x/\sqrt{L}$ 
and $\langle \vrh_k\vrhd_k\rangle_{\rm conn} =
\langle \vrh_k\vrhd_k\rangle - \langle \vrh_k\rangle\langle\vrhd_k\rangle$, 
etc. 
For a free system these formulas produce the standard distribution
functions and frequencies, while for systems out of equilibrium the above
definitions have been used successfully in \cite{Salle:2000hd,Salle:2000jb}.
A corresponding definition for fermions was introduced in 
\cite{Aarts:1999zn}, which
also mentions the relation to Wigner functions. 
Here we note in particular that
the above equations can be easily solved, with positive $\om_k$, and
we always have found $n_k$ to be positive.

The above formulas can also be applied to classical fields, with 
\be
\overline{\vr_k\vr^*_k}^{\rm conn} = n_k\, \frac{1}{\om_k},
\;\;\;\;
\overline{\pi_k\pi^*_k}^{\rm conn} = n_k\, \om_k,
\label{defnkclass}
\ee
where $\overline{\cdots}$ denotes a classical average (typically over
initial configurations).

\section{Initial conditions}

\begin{figure}
    \centering
\scalebox{1.3}{\includegraphics{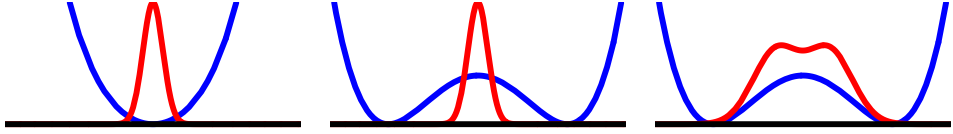}}
    \caption{The potential and sketch of
the gaussian wave function through a quench.
From left to right: $t = 0^-$, $\mu^2 > 0$; 
$t = 0^+$, $\mu^2 < 0$; a later time.}
    \label{f1}
\end{figure}

Consider a quench, a sudden change of the sign of the renormalized mass
parameter $\mu^2$ from positive to negative values at time $t=0$. This is 
supposed to mimic the effect of a rapid phase transition 
(cf.\ figure \ref{f1}).
Because the potential at $t=0^+$ is unstable, Fourier modes of the field
with wave vector $|k| <|\mu|$ are unstable and grow exponentially.
Occupation numbers grow large and non-linearities become important, which
fact has been dealt with numerically using the classical approximation. 

The question is how to start the simulation, and, can we do better
and stay within the quantum theory? Classically, the ground state at $t=0^-$
implies that the fields and canonical momenta are zero,
so nothing would happen subsequently. Quantum fluctuations are 
invoked to start the process.
%
At $t=0^-$ the system is in a single well ground state. Neglecting 
non-linearities the quantum fluctations are characterized by
\be
n_k = 0,
\;\;\;\;
\om_k = \om_k^0 = \sqrt{|\mu|^2 + k^2}.
\ee
These expectation values can be reproduced with
the classical ensemble distribution
\be
p(\vr,\pi) \propto \prod_k 
\exp\left[-\frac{\pi^*_k\pi_k + \om_k^2 \vr^*_k\vr_k}{\om_k(n_k + 1/2)}\right],
\label{gaussfielddistr}
\ee
with $n_k=0$ and $\om_k = \om^0_k$.
The type of initial conditions used in 
\cite{Khlebnikov:1996wr}-\cite{Rajantie:2000fd}
correspond to
generating configurations from this distribution and using these as initial
values for the subsequent classical evolution.
A problem with this is that the total energy density $\sum_k \om_k^0/2L$
diverges $\propto a^{-2}$ for lattice spacing $a\to 0$, and after 
equilibration the classical temperature $T$ also diverges $\propto a^{-1}$. 
Renormalization of this divergence 
is not possible out of equilibrium (in equilibrium one can attempt to do
this with temperature dependent countertems \cite{Aarts:1996qi,Aarts:1997kp}).

The Hartree approximation does not have this problem as it is formulated in the
quantum theory. However, with homogeneous initial conditions 
it has unsufficient scattering to obtain thermalization.

To improve this situation we proceed as follows. We solve the Heisenberg
equations analytically for small times, when
the modes grow exponentially $\propto \exp(t\sqrt{|\mu|^2 - |k|^2})$.
Non-linearities become important roughly when $|\mu|^2\vr^2/2 = \lm\vr^2/4$,
at the spinodal time 
$t_{\rm sp}\equiv (1/2|\mu|) \ln (2|\mu|^2/\lm)$. At some resampling time
$0< t_{\rm rs} < t_{\rm sp}$ we provide a gaussian distribution 
(\ref{gaussfielddistr}), 
for initial values $\vr_k$ and $\pi_k$ of the unstable modes,
based on the analytically calculated two-point
functions $\langle\vrhd_k\vrh_k\rangle$ and $\langle\pihd_k\pih_k\rangle$
(in the analytic solution $\langle\vrh_k\rangle
= \langle \pih_k\rangle = \langle \vrhd_k \pih_k + \pihd_k\vrh_k\rangle = 0$): 
\begin{itemize}
\item[-]
In the classical approximation, the distribution is chosen of the form
(\ref{gaussfielddistr}) with $n_k + 1/2 \to n_k$ (so dropping the 1/2), 
such that the analytically calculated
particle distribution $n_k$ is reproduced by (\ref{defnkclass}).
The high momentum modes in (\ref{gaussfielddistr})
are suppressed by $n_k$ dropping to zero 
(like $k^{-4}$ as it turns out). 
We actually keep only the unstable modes in the
initialization as for these typically $n_k \gg 1$.
The resulting energy density is finite.

\item[-]
In the Hartree ensemble approximation we 
also use the form (\ref{gaussfielddistr}) to generate
initial values for the mean fields in each Hartree realization,
again with $n_k + 1/2\to n_k$.
The mode functions are initialized as vacuum plane waves $\al\to k$,
\be
f^k_x = \frac{e^{ikx-i\om_k t}}{\sqrt{2L\om^{\rm vac}_k}}
\;\;\;\;
\om_k^{\rm vac} = \sqrt{m^2 + k^2},
\;\;\;\;
m^2=-2\mu^2,
\ee
where $m$ is the mass of the particles and the gaussian coherent states
are specified by $\hat b_k|i\rangle = 0$.
(It would perhaps have been logically
more consistent to use $\om_k$ in stead of $\om_k^{\rm vac}$,
in which case the quantum two point functions at $t=t_{\rm rs}$
would have been reproduced exactly, with the $1/2\om_k$ in (\ref{defnk})
provided by the contribution of the mode functions.) 
Since the mean fields in each realization are inhomogeneous
we expect much improved thermalization.
\end{itemize}

\section{Numerical tests}
\begin{figure}
\scalebox{1.1}{\includegraphics{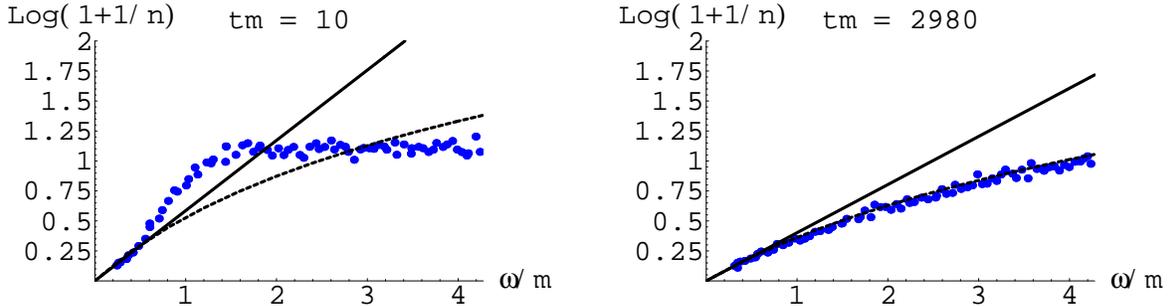}}
\caption{Particle distributions at various times using `old' initial
conditions. The straight line is a fit to the data at 
small $\om_k$. The classical distribution $n_k = T/\om_k$ with the same
temperature is also indicated (curved line).}
\label{fclassold}
\end{figure}

We used relatively weak coupling, $v^2 = m^2/2\lm = 6$, a lattice spacing
$am = 0.2$ and volume $Lm = 102.4$. 
Figure \ref{fclassold} shows the particle distribution for several times,
obtained by classical evolution using the `old' initial conditions
at $t=0^+$,
i.e.\ (\ref{gaussfielddistr}) with $n_k + 1/2 \to 1/2$ 
({\em so keeping the 1/2}) and $\om_k = \om_k^0$.
Plotted is $\ln(1+1/n_k)$ versus $\om_k$, which would give a straight
line with slope $1/T$ for a Bose-Einstein distribution 
$n_k^{-1} = \exp(\om_k/T) -1$.
One clearly sees at time $tm = 10$
that the high momentum modes, which have not been affected yet by the 
evolution,
are populated according to the vacuum fluctuations: their effective
classical particle number (from (\ref{defnk}) and (\ref{defnkclass})) is
1/2 and $\ln(1+1/0.5)\approx 1.1$. At later times the distribution approaches
the classical $T/\om_k$.

\begin{figure}
\scalebox{1.1}{\includegraphics{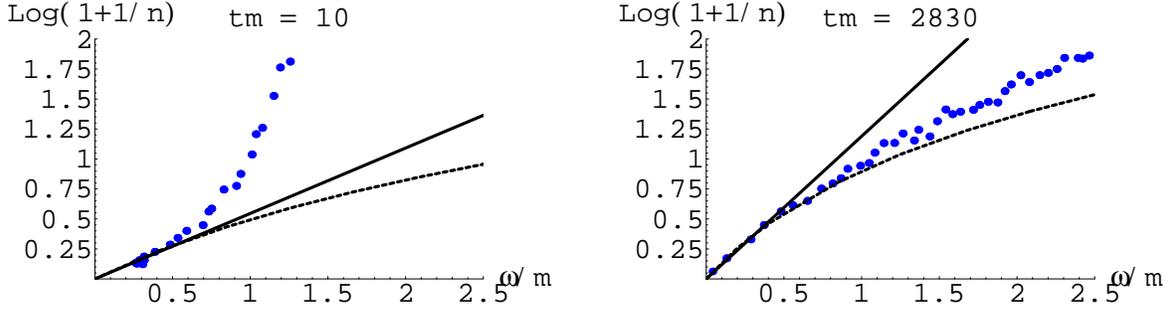}}
\caption{As in figure \protect \ref{fclassold} using the 
new initial conditions.} 
\label{fclassnew}
\end{figure}

Next we show results using the proposed `new' initial conditions for the 
classical evolution. With the parameters used here
there are then 23 unstable modes (12 with $k\ge 0$). 
The spinodal time $t_{\rm sp}m \approx 1.8$
and the resampling time was chosen to be $t_{\rm rs}m \approx 1$. 
Figure \ref{fclassnew} shows the results for this case.
We see that initially $n_k$ drops very fast as a function of $k$ 
(cf.\ $tm=10$).
At intermediate times the distribution resembles a Bose-Einstein form. The
approach to the classical distribution is much slower than in the previous
case, the flow of energy to the higher momentum modes is evidently slow.

\begin{figure}
\scalebox{0.75}{\includegraphics{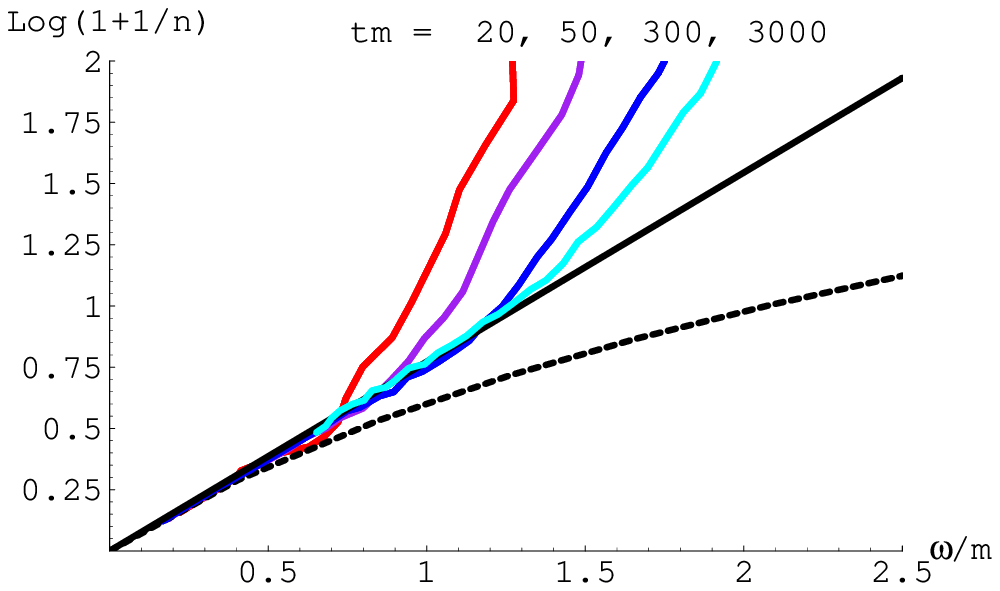}}
\scalebox{0.75}{\includegraphics{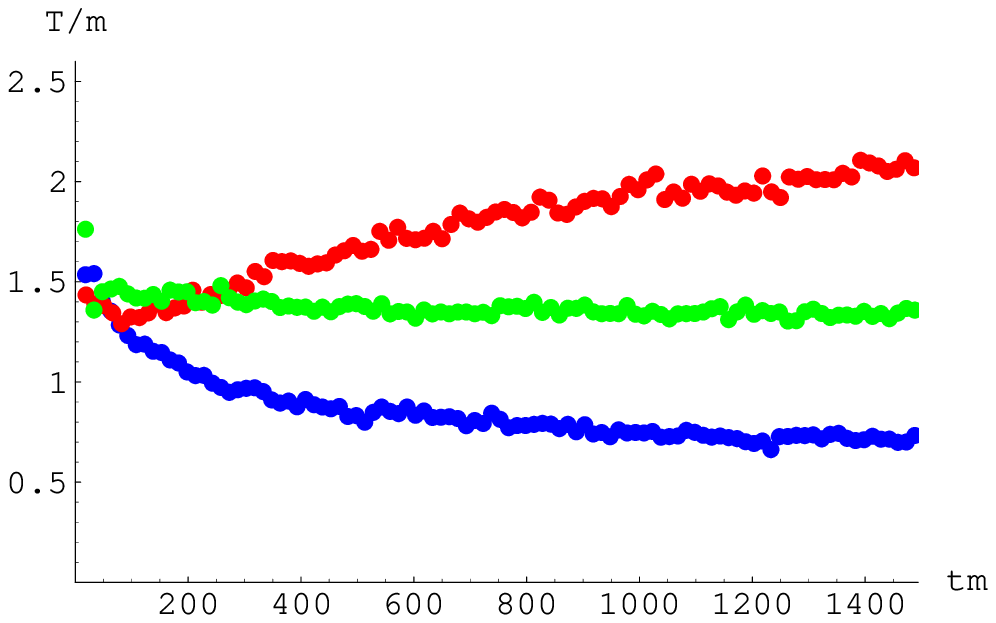}}
\caption{Left: as in figure \protect \ref{fclassold} using the 
Hartree ensemble approximation.
Right: temperatures as a function of time, for the 
classical evolution with `old' initial conditions (top), the
classical evolution with new initial conditions (bottom) and the
Hartree ensemble approximation (middle).}
\label{fquant}
\end{figure}

Results using the Hartree ensemble approximation
are shown in figure \ref{fquant}-left.
In this case we see a slow approach to a Bose-Einstein distribution, within
the same time span as in the previous cases. However,
we expect \cite{Salle:2000hd} the distribution to approach
a classical form on a much larger time scale,
typically by a factor of 100 \cite{Salle:2000jb}.

Finally, figure \ref{fquant}-right 
shows how the temperature (as determined by the
slope in figs.\ \ref{fclassold}-\ref{fquant} at low momenta) evolves in
time for the three cases.
In the first case we see the temperature gradually rising, eventually $T$ will
get of order $1/a$. In the second case $T$ drops slowly as the energy is 
being equipartitioned from the low momentum modes to all modes. For the 
Hartree ensemble case the temperature remains reasonably constant after the
initial transient. We checked that the same temperature is obtained
using a smaller resampling time.

\section{Conclusion}
The divergences in the energy density and the classical temperature are
cured by our proposal 
for obtaining initial conditions for classical simulations. It 
makes a big difference at later times. 
The Hartree ensemble approximation appears to be a significant improvement over
the classical treatments. However, this method is numerically expensive and
further tests are needed in 3+1 dimensions.

\Acknowledgements
J.S. would like to thank the organisers for an exciting conference.
This work is supported in part by FOM.

\end{document}